\documentstyle[11pt,twoside]{article}
\def\Red{}
\def\Black{}
\def\Blue{}
\def\Green{}

\newcommand{\eq}[1]{~(\ref{eq:#1})}
\newcommand{\GeV}{\,{\rm GeV}}
\newcommand{\TeV}{\,{\rm TeV}}

\newcommand{\NP}[3]{{\em Nucl. Phys. \bf #1}  (#2) #3}
\newcommand{\PRL}[3]{{\em Phys. Rev. Lett. \bf #1} (#2) #3}
\newcommand{\PL}[3]{{\em Phys. Lett. \bf #1} (#2) #3}
\newcommand{\PR}[3]{{\em Phys. Rev. \bf #1} (#2) #3}
\newcommand{\md}[1]{\langle #1 \rangle}
\newcommand{\Frac}[2]{\leavevmode\kern.1em\raise.5ex
\hbox{\the\scriptfont0 #1}
\kern-.1em/\kern-.15em\lower.25ex\hbox{\the\scriptfont0 #2}}

\makeatletter
\newcounter{alphaequation}[equation]
\def\thealphaequation{\theequation\hbox to
0.6em{\hfil\alph{alphaequation}\hfil}}
\def\eqnsystem#1{
\def\@eqnnum{{\rm (\thealphaequation)}}
\def\@@eqncr{\let\@tempa\relax \ifcase\@eqcnt \def\@tempa{& & &} \or
  \def\@tempa{& &}\or \def\@tempa{&}\fi\@tempa
  \if@eqnsw\@eqnnum\refstepcounter{alphaequation}\fi
\global\@eqnswtrue\global\@eqcnt=0\cr}
\refstepcounter{equation} \let\@currentlabel\theequation \def\@tempb{#1}
\ifx\@tempb\empty\else\label{#1}\fi
\refstepcounter{alphaequation}
\let\@currentlabel\thealphaequation
\global\@eqnswtrue\global\@eqcnt=0 \tabskip\@centering\let\\=\@eqncr
$$\halign to \displaywidth\bgroup \@eqnsel\hskip\@centering
$\displaystyle\tabskip\z@{##}$&\global\@eqcnt\@ne
\hskip2\arraycolsep\hfil${##}$\hfil& \global\@eqcnt\tw@\hskip2\arraycolsep
$\displaystyle\tabskip\z@{##}$\hfil
\tabskip\@centering&\llap{##}\tabskip\z@\cr}

\def\endeqnsystem{\@@eqncr\egroup$$\global\@ignoretrue} \makeatother

 \def\Ord{{\cal O}} 

\def\SO{{\rm SO}}

\def\circa#1{\,\raise.3ex\hbox{$#1$\kern-.75em\lower1ex\hbox{$\sim$}}\,}

\begin{document}\thispagestyle{empty}\setcounter{page}{0}
April 1996\hfill
\vbox{\hbox{\bf FT--UAM 96/14}
      \hbox{\bf hep-ph/9604417}}\\[5mm]

\vfill ~\vfill

\centerline{\huge\bf\Red Charge and color breaking minima and
\hspace{-1.945em}and }\vspace{3mm}
\centerline{\huge\bf $\!\!$constraints$\,$
on$\,$ the$\,$ MSSM$\,$ parameters}
\bigskip\bigskip\Black

\vfill

\centerline{\large\bf  Alessandro Strumia}
\bigskip
\centerline{\large\em Departamento de Fisica Te\'orica,
             m\'odulo \rm C--XI}
\centerline{\large\em Universidad Aut\'onoma de Madrid, 28049,
             Madrid, Espa\~na}
\centerline{\large and}
\centerline{\large\em INFN, Sezione di Pisa,  I-56126 Pisa, Italia}

\vfill~\vfill

\bigskip\bigskip\Blue \centerline{\large\bf Abstract}
\begin{quote}\large\indent
The MSSM potential can have unphysical minima deeper than the physically
acceptable one.
We point out that their presence is quite generic in SO(10) unification
with supergravity mediated soft terms.
However, at least for moderate values of $\tan\beta$, the physically
acceptable vacuum has a life-time longer than the universe age.
Furthermore, by discussing the evolution
of the universe in an inflationary
scenario, we show that the correct vacuum is the natural expectation.
This is not the tendency in different cases of unphysical minima.
Even in the SO(10) case,
the weak assumptions necessary for this may
prevent to use the $\tilde{\nu} h^{\rm u}$ direction
for baryogenesis {\em \'a la} Affleck-Dine.
\end{quote}\Black

\vfill~\vfill~\vfill~\vfill~
\newpage

\section{Introduction}\label{Introduction}
In the supersymmetric limit the MSSM potential possesses
some flat direction
and many almost flat directions, lifted
only by Yukawa interacions with small couplings.
In some portions of the MSSM parameter space
the soft terms give negative corrections
to the potential along such directions, so that
new other unwanted minima appear beyond the SM-like
one~\cite{A<3,L<hu,UFB}.
Typically the vacuum expectation values in the
other minima break electric
charge and/or colour and are
of order $M_Z/\lambda_f$,
where $\lambda_f$ are the Yukawa couplings
of the matter fermions~$f$.
\begin{quote}\em
What kind of restrictions on the theory
implies the possible presence of other
vacuum states beyond the SM-like one?
\end{quote}
One extremal point of view consists in barring all the
regions of parameter space where other deeper minima are present.
In this case one can derive strong constraints
on the soft terms~\cite{UFB} and, more interestingly,
some scenarios would be excluded on these grounds:
this happens, for example, in string theory when supersymmetry breaking
is dominated by the dilaton~\cite{UFB}.
Similarly, as shown in section~\ref{SO10},
the presence of deeper minima also afflicts
a SO(10) unified theory
with supergravity mediated soft terms.

At the other extremum, the weakest possible requirement on the theory
is obtained accepting the possibility that the `true' (physical)
vacuum in which we live be a `false'~\cite{Coleman}
(unstable) one\footnote{To avoid confusion we will call
`unphysical' all unacceptable vacua
different from the SM-like one.}
that some mechanism has selected among the other minima.
In this case one must at least require that the lifetime of the
unstable SM-like vacuum be bigger than the universe age.
The quantum tunneling rate
is proportional to an exponentially small factor
$\exp{\cal O}(-1/\lambda_f^2)$
which makes the SM-like minimum stable enough in all cases
except in decays towards possible minima with vacuum expectation values of
order $M_Z/\lambda_t$ (or eventually even of order
$M_Z/\lambda_b$ and $M_Z/\lambda_\tau$, if $\tan\beta$ is large).
In such a case the restrictions on the theory are
extremely weak.
In practice the only constraint derived so far
is a weak upper bound
on the top quark trilinear soft term, $A_t$~\cite{VacDec,Thermal}.
We will show that in the large $\tan\beta$ case
another well defined region of the parameter space must be excluded.

The big difference between the constraints obtained following
the two extremal attitudes
shows that a better understanding of the problem is needed.
Before excluding something due to the presence of a bad minimum
it is necessary to ensure that it is bad enough.
In other words,
we must investigate what other unacceptable
consequences can have the possible presence of
other minima beyond the SM-like one.
While their presence does not constitute
a problem for particle physics at the Fermi scale,
there can be bad {\em cosmological\/} consequences.
It may happen that
the universe evolution (as now we understand it)
can not naturally end up in the desired SM-like vacuum,
preferring instead some other unphysical minimum.

For example if the universe has passed through a hot phase
where the temperature was bigger than $M_Z$
and possibly of the order of the vacuum expectation values
of some unphysical minimum,
it is also necessary to impose that the SM-like vacuum
be stable under thermal fluctuations.
This requirement turns out to be very weak~\cite{Thermal,Riotto}.

It seems however that a cosmological scenario
compatible with particle physics
must contain a stage of inflation.
Such a scenario is most naturally realized
assuming a random initial distribution of
the order of the gravitational scale
for the various vacuum expectation values~\cite{Linde}.
If the potential during inflation would be
the usual low energy MSSM one,
then the natural end point of universe evolution
would be the `biggest' eventual unphysical minimum
with the largest vacuum expectation values.
In this case it would be natural to
avoid all the regions of the parameter space
in which another `bigger'
(and not necessarily deeper) minimum
is present.
The bounds would be even (slightly)
stronger than the strong ones.

However the same positive vacuum energy which gives rise to inflation
also produces effective soft terms of order $H$~\cite{Hsoft},
where $H$ is the Hubble constant during inflation.
In general, such new contributions to the soft terms
are not directly linked to the low energy ones.
Even when unphysical minima are generically present in the
low energy potential, the inflationary potential can be safe.
We will show that this naturally happens
in the SO(10) case discussed in section~\ref{SO10}.
As discussed in section~\ref{Inflation}, in such cases the
inflationary cosmological evolution chooses the SM-like minimum
avoiding the other ones.

\smallskip

This paper is organized as follows.
In section~\ref{BadMinima} we
briefly review the unphysical minima which turn out to be more significant
and discuss their
main characteristics.
In section~\ref{SO10} we will show that
in an SO(10) theory with supergravity mediated soft terms
the presence of a phenomenologically unacceptable deeper minimum is
quite generic.
In the large $\tan\beta$ region the resulting lifetime of the SM-like
vacuum can be smaller than the universe age.
Finally, in section~\ref{Inflation} we will discuss if
we can tolerate the presence
of other minima or if they constitute a problem for cosmology.

\section{The flat directions and the unphysical minima}\label{BadMinima}
The first example of an unphysical minima has been given in~\cite{A<3}
where it has been shown that the
(approximately) necessary and sufficient conditions
such that the RGE improved MSSM potential do not develop
a deeper minimum along the directions
\begin{eqnsystem}{sys:v1}
&&|\md{h^{\rm u}_0}| = |\md{\tilde{Q}_u}| = |\md{\tilde{u}_R}|
\sim {\cal O}(\frac{M_Z}{\lambda_u}),\\
&&|\md{h^{\rm d}_0}| = |\md{\tilde{Q}_d}| = |\md{\tilde{d}_R}|
\sim {\cal O}(\frac{M_Z}{\lambda_d}),\\
&&|\md{h^{\rm d}_0}| = |\md{\tilde{L}_e}| = |\md{\tilde{e}_R}|
\sim {\cal O}(\frac{M_Z}{\lambda_e})
\end{eqnsystem}
are respectively
\begin{eqnsystem}{sys:A<3}
A_u^2  &<& 3(\mu_{\rm u}^2 + m_{\tilde{Q}}^2 + m_{\tilde{u}_R}^2 )
\label{eq:At<3}\\
A_d^2  &<& 3(\mu_{\rm d}^2 + m_{\tilde{Q}}^2 + m_{\tilde{d}_R}^2 )\\
A_e^2  &<& 3(\mu_{\rm d}^2 + m_{\tilde{L}}^2 + m_{\tilde{e}_R}^2 )
\end{eqnsystem}
where $h^{\rm u}$ ($h^{\rm d}$) are the Higgs doublet
coupled to up quarks (down quarks and leptons),
$m_R^2$ is the soft mass for the field $R$,
$\mu_{\rm u,d}^2 = m_{h^{\rm u,d}}^2 +|\mu|^2$,
and the standard notations for the other quantities have been followed.
The generation number, which can be 1, 2 or 3, has been omitted.
Optimized versions of these directions
have then been considered in~\cite{L<hu,UFB}, obtaining
slightly more stringent constraints
in which $\mu_{\rm u,d}^2$ is replaced by $m_{h^{\rm u,d}}^2$.

Another potentially dangerous direction
have been successively considered in~\cite{L<hu}
\begin{equation}\label{eq:Lhu}
\md{h^{\rm u}} =-a^2 \frac{\mu}{\lambda_d},\qquad
\md{\tilde{Q}_d}=\md{\tilde{d}_R}=a\frac{\mu}{\lambda_d},
\qquad\hbox{and}\qquad
\md{\tilde{L}_\nu}=\frac{\mu}{\lambda_d}\cdot a\sqrt{1+a^2},
\end{equation}
where $\tilde{L}$ and the down squarks can be of any generation
and  $a$ is a numerical constant of order one
which depends on the shape of the potential.
In this case the resulting (approximate) condition
to avoid an unphysical minimum is
\begin{equation}\label{eq:L<hu}
m_{\tilde{L}}^2 + m_{h^{\rm u}}^2 > 0
\end{equation}
where $m_{h^{\rm u}}^2$ is expected to be driven negative at
some scale greater than $M_Z$ by the
top quark Yukawa coupling so that to induce the `radiative' electroweak
symmetry breaking.
Optimized versions of these directions
have been recently obtained in~\cite{UFB};
in particular the down squark may be replaced
by another slepton $\tilde{e}$ of a generation different
than the one which already appears in\eq{Lhu}.
While the approximate form of the constraint remains the same
as in eq.\eq{L<hu}, these similar minima have vacuum expectation values
of order $\mu/\lambda_e$.

\smallskip

The typical vacuum expectation value $v$ in all these minima
is of the order of the electroweak scale $M_Z$ divided by
some Yukawa coupling which may be small.
For this reason the RGE improved tree level potential
is not a good approximation:
choosing the renormalization scale at ${\cal O}(M_Z)$
the neglected full one loop potential contains terms
proportional to $\ln v/M_Z$, which may be crucial~\cite{V1loop}.
A more correct bound is obtained evaluating the running parameters
at scales $Q\sim v$.
While no significant variation is expected
in the case of the first constraint~(\ref{sys:A<3}),
the other constraint~(\ref{eq:L<hu}) becomes ineffective
when the scale $Q_0$ at which $m_{\tilde{L}}^2 + m_{h^{\rm u}}^2$
becomes negative is smaller than the typical
vacuum expectation value $\mu/\lambda_f$.
Quantum corrections are very important in this second case,
since such scale is mainly determined by the top Yukawa coupling,
which radiatively induces the electroweak breaking.
This point will be crucial in the following.
Nowadays we know that the top quark is much
heavier than what was typically assumed ten years ago
and that $\lambda_t$ at the GUT scale
is larger than about $\Frac{1}{2}$.
This implies that the scale $Q_0$ is typically
larger than $M_Z/\lambda_f$ and that the
correct evaluation of the bound~(\ref{eq:L<hu})
is still important but
no more essential.

A general analysis of the various
possibly dangerous flat directions was performed in~\cite{UFB}.
In the standard scenario of supergravity mediated soft terms
it turns out that
the two examples~(\ref{sys:A<3}) and\eq{L<hu} here reported almost include
all the others.
The first constraint gives an upper bound on
the $A$-terms, while the second one, eq.~(\ref{eq:L<hu}),
gives a $\lambda_t$-dependent upper bound on the ratio
between gaugino masses $M_{1/2}$ and scalar masses $m_0$ at the GUT scale,
that, in the case of universal soft masses at the GUT scale, is
approximately $M_{1/2}/m_0\circa{<}(0.5\div0.8)$.
This restriction can be problematic
for some scenarios where such ratio is predicted,
for example in supergravity with
dilaton dominated supersymmetry breaking~\cite{UFB}.

\begin{figure}[t]\setlength{\unitlength}{1cm}
\begin{center}
\begin{picture}(16,7)
\put(-6.3,-2.4){\includegraphics{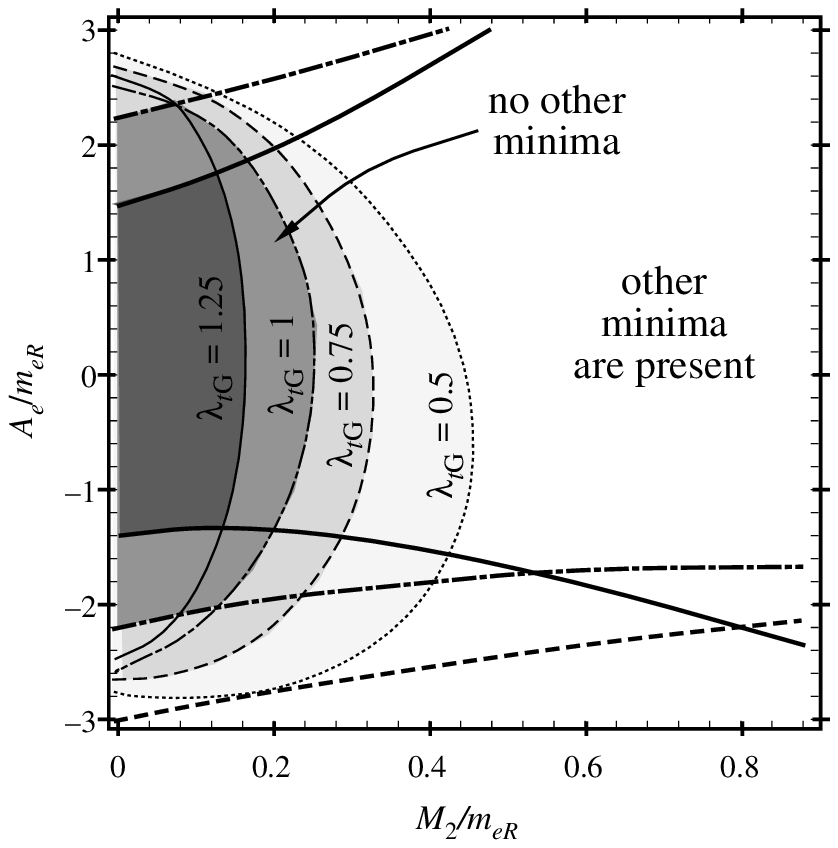}}  
\put(7.4,0){\includegraphics{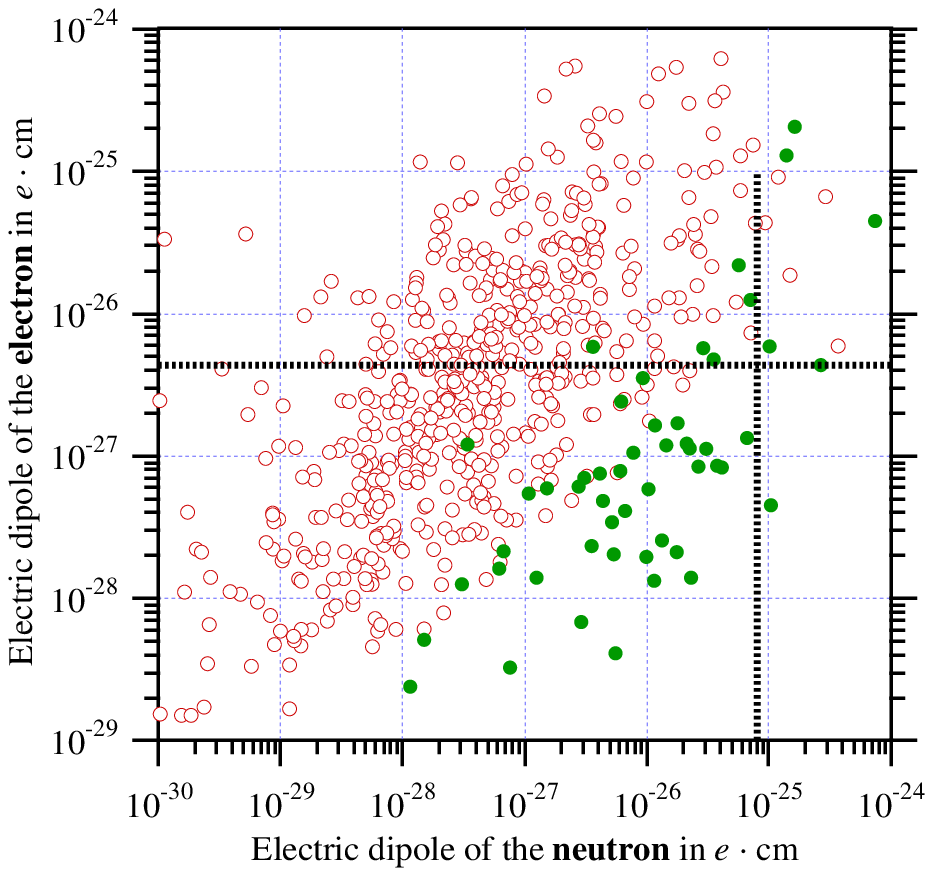}}
\end{picture}
\parbox[t]{8cm}{\caption{\em The part of all
the SO(10) parameter space\label{fig:L<hu10}
which does {\em not\/} contain an unphysical minimum
(gray area)
for different values of $\lambda_t$ at the unification scale.
The wino mass parameter, $M_2$,
the right-handed selectron mass, $m_{\tilde{e}_R}$,
and the selectron $A$-term, $A_e$, are renormalized at the weak scale.}}
\hfill
\parbox[t]{8cm}{\caption{\em Typical values
of leptonic (here $d_e$) and hadronic (here $d_N$)
signals of SO(10) unification
for different possible supersymmetric particle
spectra with (\Red$\circ$\Black) or without
(\Green$\bullet$\Black) a
deeper unphysical minimum.\label{fig:LFVgood}}}
\end{center}\end{figure}

\section{Constraints on the SO(10) parameter space}\label{SO10}
We will now show that in
a very large portion of the parameter space of
an SO(10) unified theory
with soft breaking terms mediated by minimal supergravity,
unphysical deeper minima
are present along the directions\eq{Lhu} due to
the RGE effects generated by the unified top Yukawa coupling.
These same radiative corrections to the unified soft terms
produce significant
rates for processes which violate the lepton flavour
numbers (such as $\mu\to e\gamma$)
and CP (such as $d_e$ and $d_N$)~\cite{LFV}.
For this reason the regions of parameter space
in which the unphysical minimum is not present are
the same in which the most interesting
leptonic signals of SO(10) unification
are more difficult to detect (see fig.~\ref{fig:LFVgood}).

To give a precise meaning to our computation
we will adopt the `minimal' SO(10) model
presented in ref.~\cite{LFV}
in which the only relevant
Yukawa coupling above the unification scale
is the unified top quark one and
the light Higgs doublets $h^{\rm u}$ and $h^{\rm d}$
are not unified in a single representation of SO(10),
allowing for moderate values of $\tan\beta$.
The opposite case of large $\tan\beta\sim m_t/m_b$
will be considered at the end of this section.

Let us explain why in an SO(10) GUT
the constraint~(\ref{eq:L<hu})
is stronger than in the MSSM case.
This is due to two different reasons:

\begin{itemize}
\item[i.] 
New RGE corrections are present
from the Planck scale to the GUT scale,
with the larger RGE coefficients typical of a unified theory.
This means that, for a given $\lambda_t$ value, the scale
$Q_0$ at which $m_{\tilde{L}}^2 + m_{h_{\rm u}}^2$ becomes negative
is much larger than in the MSSM.
The approximate strong form\eq{L<hu} of
the condition necessary to avoid minima in the directions\eq{Lhu}
becomes now more exact.
\item[ii.] In an SO(10) theory also the third generation
slepton doublet feels the unified Yukawa coupling of
the top and becomes lighter than the corresponding
sleptons of first and second generation.
For this reason the bound\eq{L<hu} with a slepton of third generation,
$\tilde{\nu}_\tau$,
becomes much stronger.
\end{itemize}
This explains why the bound~(\ref{eq:L<hu})
is violated in a very great part of the SO(10) parameter space,
approximately for
\begin{equation}
M_2^2 \circa{>} [0.13-0.064\lambda_t^2(M_{\rm G})] \cdot m_{\tilde{e}_R}^2
\end{equation}
where $M_2$, the wino mass parameter, and
$m_{\tilde{e}_R}$, the right-handed selectron mass, are renormalized
at the weak scale.

In the case of universal soft terms,
the precise form of this restriction is shown in figure~\ref{fig:L<hu10},
that covers all the parameter space
for any moderate $\tan\beta$ value.
Only in the shaded area inside the thin lines no other minima are present
for $\lambda_t(M_{\rm G})=0.5$ (dotted line),
$0.75$ (dashed line), $1$ (dot-dashed line) and $1.25$ (continue line).
Outside of the corresponding thick lines $m_{\tilde{\tau}}^2<0$
and the SM-like minimum is not present.
Similar results are obtained including a possible
${\rm U}(1)_X$ $D$-term contribution at the SO(10) breaking scale
or with non universal soft terms at the Planck scale.
For example, in such a case, unphysical minima
are present in all the parameter space for
$\lambda_t(M_{\rm G})\circa{>}1$
and $m_{16}^2\circa{>} 2 m_{10}^2$, where
$m_{16}$ and $m_{10}$ are the soft masses at the Planck scale for the
matter (16) and Higgs (10) SO(10) representations.

\smallskip

Let us made an aside remark.
Possible non renormalizable operators,
suppressed by inverse powers of the unification mass
or of the Planck scale,
can lift all the (almost) flat directions of the
MSSM potential
along which the unphysical minima may be present.
Anyhow, the effects of such operators are totally negligible
at vacuum expectation values
of order $M_Z/\lambda_f$, where $f$ --- in the case under examination ---
can be $d,s,b,e$ or~$\mu$.
However, a {\em light}
right handed neutrino mass $M_{\nu_R}$, around $10^{11\div13}\,{\rm GeV}$
(which in SO(10) models can be
naturally obtained as $\sim M_{\rm G}^2/M_{\rm Pl}$),
is necessary if the usual seesaw mechanism should produce
neutrino masses in the range suggested
by the MSW solution of the solar neutrino deficit. 
The superpotential then
contains the non renormalizable operators
$$
(\lambda^\nu \frac{1}{M_{\nu_R}}\lambda^\nu)
(L h^{\rm u})^2,$$
which lift the vacuum degeneracy precisely along
the very dangerous direction
under examination
$\md{\tilde{\nu}_\tau} \sim\md{ h^{\rm u}_0}$.
The corresponding dangerous minima are erased if
$m_{\nu_\tau} \circa{>} \lambda_f^2 M_Z$.
The mass of a stable $\tau$ neutrino must be smaller
than $m_{\nu_\tau}\circa{<}100\,{\rm eV}$
in order not to overclose the universe
(an unstable $\tau$ neutrino can be heavier) and
a mass around $10\,{\rm eV}$ is preferred for the (necessary?)
hot dark matter of the universe.
In this case the unphysical minima with $f=e,d$ are no more present
and those for $f=\mu,s$ only marginally affected.
In the case $f=b$ non renormalizable operators are totally irrelevant.

\medskip

We can then conclude that at least the
unphysical minimum in~(\ref{eq:Lhu}) along the
$h^{\rm u}\tilde{\nu}_\tau,~\tilde{b}_L,\tilde{b}_R$ direction
is present
in the low energy potential of an SO(10) GUT
for quite generic initial tree level values of the soft terms.
The small part of the parameter space
in which unphysical minima are not present can be characterized as follows:
\begin{itemize}
\item A {\em light chargino\/} is present in the spectrum and the
{\em squarks and the gluinos cannot be significantly heavier than
sleptons\/} (apart from a light stau), especially
if the unified top quark Yukawa coupling is near at its IR-fixed point~\cite{LFV},
$\lambda_t(M_{\rm G})\circa{>}1$.
\item With this particular spectrum,
the gluino mass RGE contribution to squark masses
can not efficiently restore their flavour universality, so that
{\em flavour and CP violating signals of SO(10) unification
in the {\em leptonic\/} sector
($\mu\to e\gamma$, $\mu\to e$ conversion, $d_e$)
are not more important than the corresponding signals
in the {\em hadronic\/} sector ($d_N$, $\varepsilon_K$, $\varepsilon'_K$,
$\Delta m_B$, $b\to s\gamma$, \ldots).}
We illustrate this point in figure~\ref{fig:LFVgood}.
We remember that lepton signals are correlated among them,
and that, at a less extent, the same happens for the
hadronic signals also~\cite{LFV}.
For this reason we have chosen
the electric dipoles of the electron, $d_e$, and
of the neutron, $d_N$, as representatives of leptonic and hadronic
signals and we have plotted in fig.~\ref{fig:LFVgood}
their predicted values in the plane $(d_N,d_e)$
for randomly chosen samples of reasonable supersymmetric spectra
and $\lambda_t$ values.
We clearly see that
an observable $d_e$ one order of magnitude larger than $d_N$,
which is a possible distinguishing feature of
SO(10) unification~\cite{dedN-SO10},
is accompanied by an unphysical minimum.
\end{itemize}
However in the next section we will argue that 
{\em unphysical minima of this kind
do not necessarily constitute a problem}
and that there is no reason to restrict the parameter space to its
small region where they are not present.

\medskip

It is now interesting to study
what happens in the large $\tan\beta$ case.
The now significant
$\lambda_\tau$ coupling below the unification scale
(together with $\lambda_{\nu_\tau}$ if $M_{\nu_R}<M_{\rm G}$)
will make the third generation sleptons even lighter
and the bound\eq{L<hu} even stronger.
The most important difference is however that
the vacuum expectation values in the
eventual unphysical minima can now be of order $M_Z/\lambda_b\sim M_Z$.
The quantum tunneling rate of the SM-like vacuum
into the unphysical vacuum would no longer contain a safe
exponentially small factor $\exp{\cal O}(-1/\lambda_f^2)$.
It may happen that
the resulting lifetime be
shorter than the universe age so that the corresponding regions
must be excluded.

\begin{figure}[t]\setlength{\unitlength}{1cm}
\begin{center}
\begin{picture}(16,8.8)
\put(-2.5,-1.5){\includegraphics{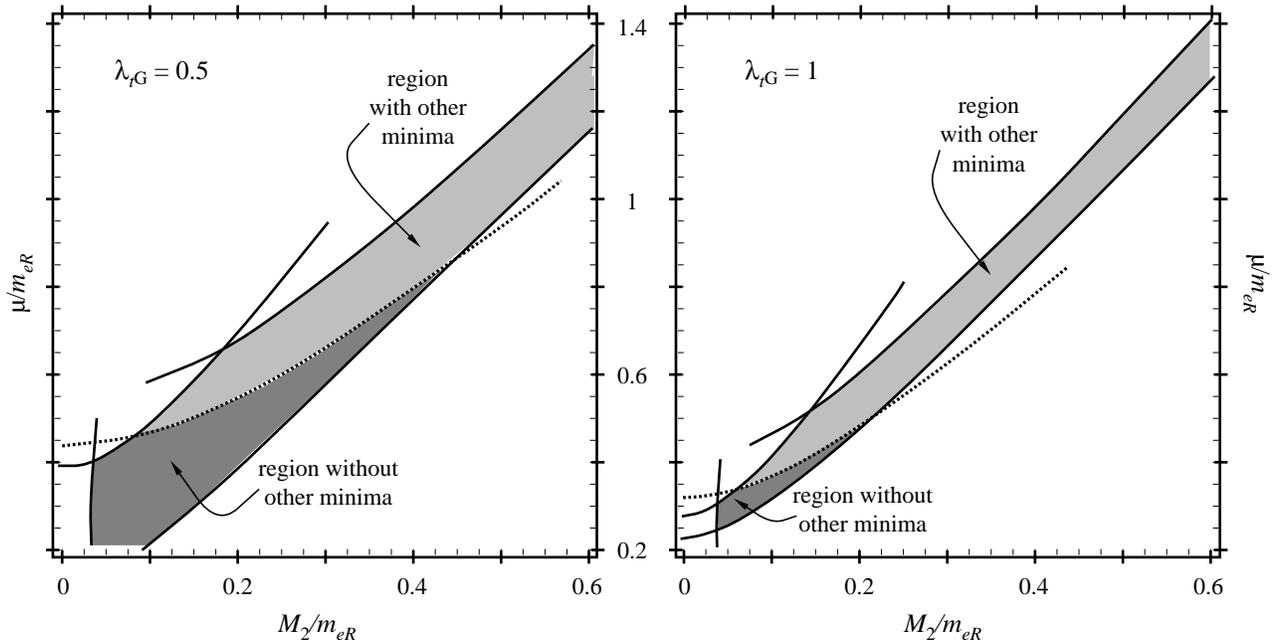}}
\end{picture}
\caption{\em  Regions of the parameter space of SO(10)
with large $\tan\beta$
without other minima beyond the SM-like one (dark gray),
and with other minima (light gray)
In the white regions the SM-like minimum is not present.
All the parameters, except $\lambda_t(M_{\rm G})$, are renormalized
at the weak scale.
\label{fig:largeTan}}
\end{center}\end{figure}

Is such an unphysical minimum again an
almost generic feature of an SO(10) GUT with large $\tan\beta$?
Of course, an appropriate numerical analysis
is necessary to delimitate the excluded regions.
It is however easy to see
that {\em the `best' part of the parameter space is safe.}
Let us remember that it is not possible to satisfy
the conditions for a
correct electroweak symmetry breaking with large $\tan\beta\sim m_t/m_b$,
\begin{equation}\label{eq:MinBigTan}
\frac{\mu  B}{\mu_{\rm u}^2 + \mu_{\rm d}^2}\approx
\frac{1}{\tan\beta}\qquad{\rm and}\qquad
\mu_{\rm u}^2 \approx -{M_Z^2 \over2},
\end{equation}
without fine tuning.
A minimum fine tuning of order $\tan\beta$
is necessary even in the `best' region~\cite{LargeTan} where
\begin{equation}\label{eq:GoodRegion}
M_i,\mu,A_f,B \sim  M_Z,\qquad\hbox{and}\qquad
m_0^2\sim \frac{M_Z^2}{\tan\beta}\sim(1\TeV)^2.
\end{equation}
Accepting to pay the price of some fine-tuning
we can exploit the full predictability of SO(10) gauge unification
considering models
where the higgs doublets
and all the third generation Yukawa couplings are unified.
In this case the resulting predictions are phenomenologically correct and
a non zero ${\rm U}(1)_X$ $D$-term
contribution $m_X^2\sim m_0^2$ is necessary
to split the $h^{\rm u}$ soft mass from the $h^{\rm d}$ one
and satisfy the minimum conditions\eq{MinBigTan}.
In this way a possible but narrow region is obtained.
The resulting particular low energy spectrum
has been explored in~\cite{LargeTan,LargeTanLFV}
and the associated rates for lepton flavour violating processes
have been computed in~\cite{LargeTanLFV},
showing that large $m_0\sim1\TeV$
sfermion masses are also preferred for compatibility
with the experimental upper bounds.

In this particular narrow
`good' region\eq{GoodRegion}
$m_{h^{\rm u}}^2 \approx \mu_{\rm u}^2 \approx -M_Z^2/2$
is much smaller than $m_{\tilde{L}_3}^2$ so that the bound\eq{L<hu}
necessary to avoid the unphysical minimum\eq{Lhu}
is already practically included in the obvious $m_{\tilde{L}_3}^2>0$
condition.

The same conclusion can not be reached if, for example,
$\mu\sim m_0$ or $m_0\sim M_Z$
which are, however, regions accessible with a worse
fine tuning~\cite{LargeTan}
and that for this reason were discarded in the
analysis in~\cite{LargeTanLFV}
making possible
more defined predictions for lepton flavour violating rates.

These qualitative considerations are confirmed
by the numerical calculation.
For example in figure~\ref{fig:largeTan} we show
the region of the $(M_2,|\mu|)$ plane
where only the SM-like minimum is present (dark gray area)
in the case of a minimal SO(10) theory
with large $\tan\beta$~\cite{LargeTanLFV}.
The figure is valid for
small values of the selectron $A$-term, $A_e\sim M_Z$,
and universal soft masses at the Planck scale
larger than the $Z$ mass as in\eq{GoodRegion}.
In the white region delimited by the various continue lines,
along which some particle becomes too light,
the SM-like minimum is not present~\cite{LargeTanLFV}.
In the remaining light gray area, which extends outside of the
preferred region\eq{GoodRegion}, the SM-like minimum is present
together with other minima since
the bound\eq{L<hu} is violated.
It is interesting that if the tunneling rate
of the SM-like vacuum were sufficiently large
one could exclude such regions on a more solid base
than fine tuning considerations.
We recall that, in standard cosmology, the probability
that the unstable SM-like vacuum has survived until today is
is given by
\begin{equation}\label{eq:tau}
p\approx\exp\big\{-(M_Z T)^4  e^{-S[\varphi^B_i]}\big\}
\end{equation}
where $T\sim 10^{10}\,{\rm yrs}$ is the universe age,
$\varphi^B_i(x)$ is the `bounce'
(a particular field configuration
which interpolates between the true and the
false vacuum)
and $S[\varphi^B_i]$ its action~\cite{Coleman}.
In our case $\varphi_i=
\{h_{\rm u},\tilde{\nu}_\tau,\tilde{b}_L,\tilde{b}_R\}$
involves more than one field
so that finding the bounce is a cumbersome numerical problem.
However, we can approximate the problem with a
single field one restricting the trajectories in field space
to the deepest direction\eq{Lhu}.
This is a good approximation for large field values,
that give a dominant and large contribution to
the bounce action, since
the potential goes down only quadratically
towards the unphysical minimum.
In such approximation the relevant Lagrangian is
\begin{equation}{\cal L}  = 2 Z(a)\, |\partial  h^{\rm u}|^2 +
\big\{m_2^2 |h^{\rm u}|^2 -
\frac{|\mu|}{\lambda_b} m_3^2 |h^{\rm u}|\big\}
\end{equation}
where $m_2^2\equiv |m_{h^{\rm u}}^2 + m_{\tilde{\nu}_\tau}^2|$,
$m_3^2\equiv  m_{\tilde{\nu}_\tau}^2+m_{\tilde{b}_L}^2+m_{\tilde{b}_R}^2$,
and
\begin{equation}\label{eq:a}
a =\left|\frac{h^{\rm u}}{\mu/\lambda_b}\right| \ge 0,\qquad 
Z(a) = \frac{8 a^2 + 10 a + 3}{8a(a+1)}.
\end{equation}
At `large' field values $a\circa{>}1$
where this Lagrangian is realistic we can further
approximate $Z\approx 1$.
Note that we are now going to compute the tunneling rate between two minima
using an approximated potential that does not have any minimum.
Since this might seem suspect, 
it is better to add some word of comment.
The {\em unphysical minimum\/} appears at the (high) scale at which
the scale-dependent $m_{h^{\rm u}}^2 + m_{\tilde{\nu}_\tau}^2$
term becomes positive.
We can however neglect this scale dependence,
since the behavior of the potential beyond the `escape' point,
$\varphi_i^B(0)$ that in our case if of $\Ord(\TeV)$,
is irrelevant~\cite{Thermal},
and, in fact,
the potential could even be unbounded from below
along the direction\eq{Lhu}.
About the {\em unstable minimum\/} at $a\approx 0$,
we only need to proceed carefully and
consider the potential as the limit of
a `conventionally shaped' one, different from our
approximated form only for $a\to 0$.
In this sense we can work with a potential without minima.
While being a bit crazy, our approximation
constitutes a great simplification, since now
the bounce action depends only on {\em one}
dimensionless ratio $[\mu m_3^2]^2/[m_2^2]^3$.
Moreover, expressing the Lagrangian in terms of
a $h^{\rm u}$ field normalized in units of
$\mu/\lambda_b$ as in eq.\eq{a},
we can see that,
due to the particular form of the approximated potential,
$S[\varphi^B_i]\propto (\mu/\lambda_b)^2$.
These considerations fix the bounce action to be
\begin{equation}\label{eq:S}
S[\varphi^B_i]  \approx c\cdot \frac{2\pi^2}{\lambda_b^2}
\frac{\mu^2(m_{\tilde{\nu}_\tau}^2+m_{\tilde{b}_L}^2+m_{\tilde{b}_R}^2)^2}
{|m_{\tilde{\nu}_\tau}^2+ m_{h^{\rm u}}^2|^3}.
\end{equation}
The dimensionless
proportionality constant $c$ can be easily computed
since, under our approximations,
it is also possible to find analytically the bounce
$h^{{\rm u}B}$
in terms of the Bessel function $J_1$,
$$h^{{\rm u}B}(r) = \frac{|\mu| m_3^2}{2 \lambda_b m_2^2}
\times\left\{\begin{array}{ll} 
b + (1-b)j_1(m_2 r)& 
\hbox{for $0\le r\le r_*$}\\
0 & \hbox{for $r\ge r_*$}\end{array}\right. ,\qquad\qquad$$
where $r$ is the Euclidean 4-radius,
$j_1(x)\equiv 2J_1(x)/x$,
$m_2r_*\approx 5.14$ is the position of its first minimum, and
$b\approx 1/8.56$ has been chosen in such a way that
$h^{{\rm u}B}(r_*)=0$.
The relatively large value of $r_*$,
due to the slow quadratic decrease of the potential
at large $h^{\rm u}$,
give rise to a correspondingly
large proportionality
constant in eq.\eq{S}, $c\approx 90$.


Let us now apply these results to the minimal SO(10) case
with universal soft terms at the Planck scale and large $\tan\beta$,
so that $\lambda_b\approx 0.9$.
In this model there are peculiar correlations
between the soft parameters~\cite{LargeTanLFV}.
In particular there exists no deep interior area where the
bound\eq{L<hu} is sufficiently strongly violated,
as shown in figure~\ref{fig:largeTan}.
For this reason, in all the allowed
parameter space, $S[\varphi^B_i]\approx 10^{4\div5}$
is quite large,
so that {\em the lifetime of the SM-like minimum is always much larger
than the universe age}.
The fact that the bounce action is always much larger
than the limiting value
ensures that the quality of our approximation
is much better than what had been sufficient.
For more general but less interesting supersymmetric
models with large $\tan\beta$ --- for example not imposing
exact SO(10) relations at the unification scale ---
the parameter space is more various and
the life-time of the SM-like vacuum can be shorter than the universe age.

\section{Cosmological evolution and unphysical minima}\label{Inflation}
We now come back to the question raised in the
introduction.
What are the consequences of the possible
presence of phenomenologically unacceptable
minima other than the SM-like one?
For example, is it necessary to
restrict an SO(10) theory to the narrow
part of its parameter space where unphysical minima are not present?

As discussed in the introduction there can be cosmological problems:
if the low energy potential possesses more than
one minimum
it may be impossible, or unnatural,
that the SM-like minimum be the actually populated one.
Implementing this condition requires considering
the universe evolution, so that
the answer is not only a matter
of particle physics at the Fermi scale.
While to obtain precise results would be necessary to choose some
particular well defined
cosmological model, the general ideas
at the basis of our present understanding of universe
evolution are, for example, sufficient to clarify the situation in the
SO(10) case presented in the previous section.
For this reason, we will not adopt any particular cosmological model
but rather we will base our discussion only on its
most well established features.
We can summarize them as follows.
A period of inflation
seems to be an essential feature
of any consistent cosmological model.
After inflation the inflaton field decays
in a time of order $ \Gamma_I^{-1}$ giving rise to
the standard `hot big-bang' scenario
with a reheating temperature $T_R\sim (\Gamma_I M_{\rm Pl})^{1/2}$.
The natural way of obtaining a sufficient amount of inflation
consists in assuming random out of equilibrium initial values
of the order of the Planck scale for the various fields.
In such a case the non zero modes of the fields
are rapidly red-shifted away and the
various fields begin to move towards a minimum.

When more than one minimum is present,
{\em the most natural
final vacuum state is the one with the largest
vacuum expectation value.}
For example, in the case of a single field whose potential possesses
two minima at different scales in field space
$v\sim M_Z$ and $v'\sim v/\lambda_f$
the relative probability is at least $p/p'\sim\lambda_f$,
even in the case where
the depths are comparable.
This shows that the unphysical minima with the smallest $\lambda_f$
are the more dangerous ones,
even if not deeper than the SM-like one.
Here we are assuming that the unknown mechanism
responsible of the small present value of
the cosmological constant
does not treat the SM-like minimum as special.

We thus reach the conclusion that the bounds
can be even (slightly) stronger than in the extremal case
of excluding only all deeper unphysical minima.
However such conditions
can not be directly applied to the low energy potential,
since, during inflation,
the potential along the flat directions
is expected to be significantly different
from the low energy one~\cite{Hsoft}.
The reason is that
the same positive vacuum energy density $V=|F_I|^2+\cdots$
which give rise to inflation
with a Hubble constant $H^2\sim V/M_{\rm Pl}^2$
also produces supergravity mediated
soft terms of order~$H$~\cite{Hsoft}.
If the inflaton field $\varphi_I$
were more directly coupled to the SM fields
than the usual `hidden' sector
responsible of low energy supersymmetry breaking,
the resulting effective soft terms
would be even bigger.
For example $\varphi_I$ may have a Yukawa coupling
to some charged field with mass
of the order of the GUT scale~\cite{Hsoft}.
In the opposite case, for example in no-scale models,
it is also possible
that the inflationary soft terms
are zero at tree level~\cite{noHsoft}.
In general the inflationary contributions to the soft terms
are not directly linked to the low energy ones
neither in the supergravity case.
In fact, a vacuum expectation value
of the inflaton away from the minimum and of the order of the
Planck scale, $\md{\varphi_I}\sim M_{\rm Pl}$,
can distort the K\"ahler potential.

To summarize we can say that,
while the necessary bounds are possibly
very strong, they do not depend
only on physics at the Fermi scale and
their computation requires a knowledge
of physics at the Planck scale.
The well established features of
cosmological evolution alone are not sufficient
to determine if a particular unphysical minimum
is present even in the inflationary potential
so that to decide if it
has untolerable consequences which force to exclude its presence.
On the contrary, the identification of the inflaton with
some particular field of known behavior
(such as the dilaton) would allow to precise
the bounds on the low energy theory.

It is interesting that we can obtain a definite
answer in the $\SO(10)$ case where
an unphysical minimum is present in the {\em low\/} energy potential
for quite generic
values of the soft terms
due to combined effect of $\lambda_t$-induced renormalization
corrections at all scales from $M_{\rm Pl}$ down to $M_Z$.
We now show that unphysical minima
do not afflict the potential during inflation
with soft terms of order $H$.
We recall that $H\sim 10^{14}\GeV$ is
the preferred value for which the
quantum de$\,$Sitter fluctuations of the inflaton give rise to
density perturbation which produce
the observed amount of large scale inhomogeneities.
In the case where the dominant contribution
to inflationary soft terms is transmitted
by some field at or below the GUT scale,
the {\em unified\/} top quark Yukawa coupling can
no longer affect the potential
distorting it to a dangerous form.
Even in the simplest case of supergravity mediated soft terms,
the quantum corrections
at scales greater than $H$ alone, are not sufficient
for creating unphysical minima.
We can conclude that, in SO(10) unification (but, of course, also
in more general models) quantum corrections do not
generate unphysical minima along the direction\eq{Lhu}
in the inflationary potential,
at least until $H$ is sufficiently large.
Of course unphysical minima can reappear when inflation is over.

\medskip

In a generic case it is possible that, just as in the SO(10) case,
the unphysical minimum of the low energy potential is not present in the
inflationary potential.
We now show that in such cases the cosmological problems disappear.
In fact, if the potential during inflation does not contain other minima,
when inflation is over, the various fields $\varphi$
have already efficiently rolled towards zero~\cite{AD2},
ending up, ultimately, in the SM-like minimum.
It is again crucial that the inflationary soft terms be of the order of
the inflationary Hubble constant,
so that the `forcing' $V'$ term in the inflationary field equations
\begin{equation}
\ddot{\varphi} + 3H\dot \varphi+ V'(\varphi)=0,\qquad
V(\varphi)={\cal O}(H^2) |\varphi|^2 +\cdots+
\frac{|\varphi|^{n+4}}{M^n}
\end{equation}
is at least of the same order of the `damping' term.

Let us assume that the inflationary soft masses squared
${\cal O}(H^2)$ be positive.
This is indeed what happens for a
minimal form of the K\"ahler potential~\cite{Hsoft}.
In this case the fields evolve as
$\varphi(t)\sim \varphi_0 \cdot e^{-Ht}\cos Ht$, and,
even starting from Planck scale values
$\varphi_0 \sim M_{\rm Pl}$,
at the end of inflation $\varphi< M_Z$
provided that the number of $e$-foldings be sufficiently big.
This may be difficult to obtain for a given initial
starting condition, but it is also exactly what
is necessary if inflation should dilute
unwanted species and produce the necessary
homogeneity and isotropy.
Moreover this is naturally obtained in the
`chaotic' inflationary scenario we have assumed,
where the regions in which the most favorable
starting conditions are satisfied
are the ones which expand experiencing a huge amount of inflation.

We have shown that inflationary cosmology avoids
unphysical minima present only
in the low energy MSSM potential.
If they were instead present also in the
inflationary potential the `largest' minimum would be preferred
as the true vacuum of the universe.

It is also possible that
another minimum is present in the inflationary potential
but not in the low energy MSSM potential.
This happens if one soft mass squared ${\cal O}(H^2)$ is negative and if,
for example, the usual
quantum corrections at energies $E\ll H$ stabilize
the potential at low field values, $\varphi\circa{>}M_Z$.
In this case the fields roll in the moving minimum
$\md{\varphi}(t)\sim [H(t) M^{n+1}]^{1/(n+2)}$ following it~\cite{AD2}
and then relaxing towards the SM-like minimum
when the Hubble constant $H$ and the temperature $T$
became small enough that the other minima disappear
and only the SM-like minimum is present.
This is not a problematic situation.
On the contrary, it has been shown~\cite{AD2,AD} that
if lepton (or baryon) number and CP are broken in the inflationary minimum
a baryon asymmetry is produced,
which may be easily the source of the observed one.
Since the viability of alternative mechanisms for generating it
at the electroweak scale is not ensured,
another minimum in the inflationary potential is welcome.
We only need to point out that if a corresponding
unphysical minimum, or one sufficiently coupled to it in the inflationary
field equations, is present also in
the low energy MSSM potential, then one
has to worry that the SM-like minimum might not be the populated one.

\medskip

Having discussed how inflation can naturally select the SM-like minimum
as the physical one even if bigger
minima are present in the MSSM potential,
before concluding that unphysical minima are not dangerous in these cases,
we must first ensure that the universe remains in the SM-like vacuum
until the present epoch.
This might fail due to the following effects:
\begin{itemize}
\item
{\em Quantum fluctuations\/} of the MSSM fields in
de$\,$Sitter inflationary universe with
$\md{\delta\varphi^2}\sim H^2$
and correlation length of order $H^{-1}$~\cite{deSfluc}
are not sufficient to populate the
unphysical minima with large vacuum expectation values $H/\lambda_f\gg H$.
At the end of inflation, when $H(t)\sim M_Z$, such fluctuations
could be a problem {\em if}
an unphysical minima with $\md{\varphi}\sim M_Z$
is present, as may happen if the trilinear soft term $A_t$
of the top quark exceeds the bound\eq{At<3}.
\item
In the subsequent `hot big-bang'
phase after inflation, where the reheating temperature $T_R$
is expected to be bigger than $M_Z$
and possibly of the order
of the unphysical minima vacuum expectation values,
it is also necessary to ensure that the `false' SM-like vacuum
be stable under {\em thermal fluctuations}.
This requirement turns out to be very weak~\cite{Thermal,Riotto}.
It is also possible that
high temperature corrections stabilize sufficiently
the potential, for example along the direction\eq{Lhu},
erasing the corresponding unphysical minima
during the high temperature phase after inflation.
In this case, excluding a
small part of the MSSM parameter space
with $m_{\tilde{t}_R}^2<0$~\cite{Thermal},
the symmetric vacuum will do a phase transition towards
the SM-like one
when the temperature cools down below the electroweak scale.
\item
Finally, the unstable SM-like vacuum must not
undergo {\em quantum tunneling\/} towards a deeper unphysical minimum
in the subsequent $10^{10}\,{\rm years}$.
The relative probability is $1-p$, with $p$ given in eq.\eq{tau} in terms
of the `bounce' action~\cite{Coleman}
$S[\varphi^B]={\cal O}(2\pi^2/\lambda_f^2)$.
For moderate values of $\tan\beta$ only minima
with vacuum expectation values of order $M_Z/\lambda_t$
can be dangerous,
giving rise to a weak upper bound on the trilinear soft term $A_t$
of the top quark~\cite{Thermal}.
\end{itemize}
In the SO(10) case,
where unphysical minima with vacuum expectation values
of order $M_Z/\lambda_b$ or larger
are present in the low energy MSSM potential,
the SM-like minimum suffers no instability
in the moderate $\tan\beta$ region.
On the contrary quantum tunneling effects
can be dangerous if $\tan\beta$ is large,
excluding some region of the parameter space,
as discussed in section~\ref{SO10}.

\section{Conclusions}
The MSSM potential can have other minima deeper than the SM-like one.
Their presence can be quite generic in some scenarios;
for example in section~\ref{SO10} we have shown that, in
SO(10) unification with supergravity mediated soft terms,
unphysical minima with vacuum expectation values of order
$M_Z/\lambda_b$ or larger
are not present only
in small and well defined regions of the parameter space
with a peculiar phenomenology.
In the large $\tan\beta$ case
no unphysical minima are present in the preferred part\eq{GoodRegion}
of the parameter space which is
accessible with a minimum fine tuning of order $\tan\beta\sim m_t/m_b$.
Due to $\lambda_b\sim 1$, in some region
the quantum tunneling rate of the SM-like minimum can be too large
and it is clear that such regions must be excluded.
On the contrary,
if $\tan\beta$ is small, the lifetime of the
SM-like vacuum is exponentially larger than the
universe age.

In general, it is not clear whether the presence
of other deeper minima signals a problem of the theory.
For this reason we have studied if unphysical minima have
other unacceptable consequences which force
to exclude their presence, and consequently
the scenarios that predict them.
From the point of view of particle physics,
the presence of other minima
does not affect the physics in the SM-like minimum.
It can instead be a problem for cosmology.
However, once we have accepted that for some reason
the SM-like minimum is the populated one,
the conditions necessary to ensure sufficient stability are very weak.

Stronger bounds could be obtained
studying why the SM-like vacuum should be the populated one.
In particular this question can be attached
in an inflationary scenario, where
chaotic vacuum expectation values of the order of the Planck scale
are suggested as starting conditions
in order to obtain the necessary amount of inflation.
In this case, we have seen that
if more than one vacuum is present in
the {\em inflationary} potential, the subsequent evolutions
naturally
prefers the
minimum with the largest vacuum expectation value,
which in many interesting cases is not the SM-like one.
However, even if strong bounds must be imposed on the
parameters of the inflationary potential,
we presently can not use them to constrain the low energy physics.
The reason is that new dominant contributions to the soft terms
of the order of the inflationary Hubble constant
are present during inflation
and we can not compute them without a well defined model of inflation.
In particular it turns out that if no other minimum
is present during inflation
(or at least during a sufficiently long stage of it),
then the SM-like minimum,
being the basin of the zero point in field space,
is naturally selected
even when the low energy potential contains other minima.
We have shown that this is exactly
what happens in the $\SO(10)$ case
since the effect which generates the unphysical minima at low energy
is not operative during inflation.
In this case the SM-like minimum will be selected,
provided that the tree level inflationary potential is safe.
The opposite assumption is necessary if baryogenesis
should be produced by the Affleck-Dine mechanism~\cite{AD},
at least along the `ideal' $\tilde{\nu}h^{\rm u}$ direction~\cite{AD2}.

\subsection*{Acknowledgments}
The author would like to thank R. Barbieri, G. Dvali, H.B. Kim and
C. Mu\~noz for helpful discussions.


\frenchspacing
\begin{thebibliography}{nn}
\bibitem{A<3} J.M. Frere, D.R.T. Jones and S. Raby, \NP{B222}{1983}{11}.
\bibitem{L<hu} H. Komatsu, \PL{B215}{1988}{323}.
\bibitem{UFB} J.A. Casas, A. Lleyda and C. Mu\~noz, \NP{B471}{1996}{3}
and  \PL{B380}{1996}{59}.
\bibitem{Coleman} S. Coleman, \PR{D15}{1977}{2929};\\
S. Coleman and C. Callan, \PR{D16}{1977}{1762};\\
A. Kusenko, \PL{B358}{1995}{47}.
\bibitem{VacDec} M. Claudson, L. Hall and I. Hinchliffe,
\NP{B228}{1983}{501}.
\bibitem{Thermal}
A. Kusenko, P. Langacker and G. Segre, hep-ph/9602414.
\bibitem{Riotto}A. Riotto and E. Roulet, \PL{B377}{1996}{60}.
\bibitem{Linde} A.D. Linde, {\em ``Inflation and quantum cosmology''},
Academic Press, Inc. 1990.
\bibitem{Hsoft}
M. Dine, W. Fishler and D. Nemeschansky, \PL{B136}{1984}{169};\\
O. Bertolami and G. Ross, \PL{B183}{1987}{163};\\
G. Dvali, \PL{B355}{1995}{78} and hep-ph/9503259;\\
M. Dine, L. Randall and S. Thomas, \PRL{75}{1995}{398}.
\bibitem{V1loop} G. Gamberini, G. Ridolfi and F. Zwirner,
\NP{B331}{1990}{331}.
\bibitem{LFV} R. Barbieri and L. Hall, \PL{B338}{1994}{212};\\
S. Dimopoulos and L. Hall, \PL{B344}{1995}{185};\\
R. Barbieri, L. Hall and A. Strumia,
\NP{B445}{1995}{219} and \NP{B449}{1995}{437}.
\bibitem{dedN-SO10} R. Barbieri, A. Romanino and A. Strumia,
\PL{B369}{1996}{283}.
\bibitem{LargeTan} L. Hall, R. Rattazzi and U. Sarid,
\PR{D50}{1994}{7048};\\
R. Rattazzi and U. Sarid, \PR{D53}{1996}{1553}.
\bibitem{LargeTanLFV}
P. Ciafaloni, A. Romanino and A. Strumia, \NP{B458}{1996}{3}.
\bibitem{noHsoft}
M. Gaillard, H. Murayama and K.A. Olive, \PL{B355}{1995}{71}.
\bibitem{AD2} M. Dine, L. Randall and S. Thomas, \NP{B458}{1996}{291}.
\bibitem{AD} I. Affleck and M. Dine, \NP{B249}{1985}{361}. 
\bibitem{deSfluc}
T. Bunch and P. Davies, {\em Proc. R. Soc. \bf A360 \rm (1978) 117};\\
A. Linde, \PL{B116}{1982}{335} and \PL{B131}{1983}{330};\\
A.A. Starobinskii, \PL{B117}{1982}{175}.
\end{thebibliography}
\end{document}